\documentclass[twocolumn,showpacs]{revtex4}
\usepackage{amsmath,amssymb,dcolumn}
\usepackage[dvips]{graphics}
\usepackage{graphicx}
\def\imo{i}
\begin{document}
\title{Quasi-normal modes of Brane-Localised Standard Model Fields in Gauss-Bonnet theory}
\author{A. Zhidenko}\email{zhidenko@fma.if.usp.br}
\affiliation{Instituto de F\'{\i}sica, Universidade de S\~{a}o Paulo \\
C.P. 66318, 05315-970, S\~{a}o Paulo-SP, Brazil}

\begin{abstract}
We study the massless scalar, Dirac and electro-magnetic fields propagating on a $4D$-brane, which is embedded in higher dimensional Gauss-Bonnet space-time. We calculate, in time-domain, the fundamental quasi-normal modes (QNMs) of a spherically symmetric black hole for such fields. Using WKB approximation we study QNMs in the large multipole limit. We observe also a universal behavior, independent on a field and value of the Gauss-Bonnet parameter, at asymptotically late time.
\end{abstract}

\maketitle

\section{Introduction}
Last years, higher dimensional gravity have been studied in the context of solving of the hierarchy problem \cite{ADD}. One of the possible scenarios is that the Standard Model particles (scalars, fermions and gauge bosons) are restricted to live on the $3+1$-brane, which is embedded in the higher-dimensional 'bulk', while the gravitons can propagate also in the bulk. We could detect this phenomenon by creating mini black holes at particle collisions in Large Hadron Collider (LHC), probably at energies of $\sim1TeV$ \cite{creation}. The creation of these black holes, and thus the existence of additional dimensions, can be confirmed by observing secondary effects, such as Hawking radiation or quasi-normal ringing \cite{Kanti:2004nr}.

Higher dimensional quantum gravity implies corrections to classical general relativity. The dominant order correction to the Lagrangian is called the Gauss-Bonnet term. This term is squared in curvature and vanishes for $D=4$. Recent years, black holes in the Einstein-Gauss-Bonnet theory \cite{Boulware:1985wk} have been extensively studied \cite{GBgeneral}. Such black holes were found to be unstable for $D=5,6$, when the Gauss-Bonnet parameter $\alpha$, measured in units of the horizon radius, is large \cite{GBunstable}. This fact suggests that the first correction to the Lagrangian is not enough to describe very small black holes properly. Yet, we are able to study black holes if $\alpha$ is small. The Gauss-Bonnet approximation provides also qualitative information about the influence of the Gauss-Bonnet term, when $\alpha$ is large.

One of the most important properties of any black hole is its quasi-normal ringing. Being independent on an initial perturbation, the quasi-normal oscillations are functions of the black hole parameters. These oscillations decay exponentially with time and are conveniently described by a set of complex frequencies. Their real parts are the actual oscillation frequencies, while the imaginary parts correspond to the damping rates. The quasi-normal spectrum observation allows not only to detect black holes, but also to evaluate their parameters. That is why the quasi-normal ringing was extensively studied in the context of higher dimensional theories \cite{QNMshigherD}. The gravitational perturbation and thermodynamical properties of black hole solutions in some of the brane world scenarios were studied in \cite{Abdalla:2006qj}. The quasi-normal modes of Gauss-Bonnet black holes were studied for the test scalar field \cite{Abdalla:2005hu,GBscalar} and for the gravitational perturbations \cite{Konoplya:2008ix}.

While the quasi-normal ringing of the brane-localised Standard Model fields, has been studied for Reissner-Nordstr\"om, Schwarzszchild-(anti)de Sitter \cite{Kanti:2005xa} and Kerr black holes \cite{Kanti:2006ua}, the influence of the Gauss-Bonnet term on the quasi-normal spectrum is still unknown, though it is significant for the black holes created by particle collisions \cite{Rychkov:2004sf}. The aim of our work is to fill this gap. We perform a complete numerical analysis of the evolution of the brane-localised Standard Model fields near a $D$-dimensional Gauss-Bonnet black hole with $D = 5-11$ in order to study their quasi-normal ringing and late-time tails.

The paper organized as follows. Sec. \ref{sec:basic_eq} introduces the metric for Gauss-Bonnet black holes and the wave-like equations for the brane-localised fields. Sec. \ref{sec:timedomain} describes the method of time domain integration used here and presents the numerical results. In Sec. \ref{sec:lmlimit} we study the quasi-normal spectrum in the large multipole limit. Finally, in Sec. \ref{sec:conclusion} we discuss the obtained results.

\begin{widetext}
\section{Basic equations}\label{sec:basic_eq}

The Lagrangian of the Einstein-Gauss-Bonnet action is
\begin{equation}
I = \frac{1}{16 \pi G_{D}} \int{d^{D} x \sqrt{-g} R} + \alpha^\prime
\int d^{D} x \sqrt{-g} (R_{abcd} R^{abcd}- 4 R_{cd}R^{cd} + R^2).
\end{equation}
Here $\alpha^\prime$ is a positive coupling constant.

The spherically symmetric stationary solution of corresponding Einstein equations is a Gauss-Bonnet black hole, which is given by the metric
\begin{equation}\label{metric}
ds^2=f(r)dt^2-\frac{dr^2}{f(r)}-r^2d\Omega_{D-2}^2,
\end{equation}
$$f(r)=1+\frac{r^2}{\alpha(D-3)(D-4)}\left(1-\sqrt{1+\frac{4\alpha(D-3)(D-4)\mu}{(D-2)r^{D-1}}}\right),$$
where $\mu$ is proportional to the black hole mass, $\alpha = 16 \pi G_{D} \alpha^\prime$.

In order to measure all the quantities in terms of the black hole horizon radius $r_0$, we parameterize
the black hole mass as
\begin{equation}
\mu=\frac{(D-2)r_0^{D-3}}{4}\left(2+\frac{\alpha(D-3)(D-4)}{r_0^2}\right).
\end{equation}

Since we study small black holes within large extra dimensions scenarios the effective metric background for the Standard Model field propagation on a $4D$-brane is given by the projection of the higher-dimensional one onto the brane by fixing the values of the additional angular coordinates that describe the $D-4$ extra space-like dimensions \cite{Kanti&March-Russell}. Thus, we assume that the metric on the brane has the form
\begin{equation}\label{inducedmetric}
ds^2=f(r)dt^2-\frac{dr^2}{f(r)}-r^2d\sigma^2, \qquad d\sigma^2 = d\theta^2+\sin^2\theta d\phi^2.
\end{equation}

The massless scalar field $\Phi$, the massless Dirac field $\Psi$ and the Maxwell field $A_\mu$ satisfy the equations of motion in curved space-time, described by the metric (\ref{inducedmetric}):
\begin{eqnarray}
\partial_\nu(\sqrt{-g}g^{\mu\nu}\partial_\mu\Phi)&=&0, \\
\gamma^ae_a^{~\mu}(\partial_\mu+\Gamma_\mu)\Psi&=&0, \\
\partial_\nu\left(g^{\tau\mu}g^{\sigma\mu}\sqrt{-g}(\partial_\sigma A_\tau-\partial_\tau A_\sigma)\right)&=&0.
\end{eqnarray}

After the separation of radial and angular variables the radial part of the fields equations of motion are reduced to the wave-like equations of the form
\begin{equation}\label{wave-like}
\left(\frac{\partial^2}{\partial t^2}-\frac{\partial^2}{\partial r_\star^2}+V(r)\right)R(t,r)=0,
\end{equation}
where $\displaystyle dr_\star=\frac{dr}{f(r)}$ is the tortoise coordinate.

The effective potentials are given by
\begin{equation}\label{scalar-potential}
V_s(r)=f(r)\left(\frac{l(l+1)}{r^2}+\frac{f'(r)}{r}\right), \qquad l = 0, 1, 2\ldots
\end{equation}
\begin{equation}\label{Dirac-potential}
V_{D\pm}=f(r)\frac{\kappa_{\pm}^2}{r^2}\pm \frac{d}{dr_\star}\frac{\kappa_{\pm}\sqrt{f(r)}}{r}, \qquad \kappa_{\pm}=1,2,3\ldots
\end{equation}
\begin{equation}\label{gauge-potential}
V_g(r)=f(r)\frac{l(l+1)}{r^2}, \qquad l = 1, 2, 3\ldots
\end{equation}
for the massless scalar, massless Dirac and Maxwell fields respectively. Indices $l$ and $\kappa_\pm$ parameterize the angular separation constant. Since the effective potentials $V_{D\pm}$ are isospectral we consider only the ``plus'' case \cite{Zhidenko:2003wq}.

\section{The evolution of perturbations in time domain}\label{sec:timedomain}

\begin{table}
\caption{Fundamental quasi-normal modes of the Gauss-Bonnet black hole for the test scalar field localised on the $4D$-brane ($l=1$). All quantities are measured in units of the radius of horizon. The first line ($\alpha=0$) data are taken from \cite{Kanti:2006ua}.}\label{tbl.s=0.l=1}
\begin{tabular}{|c|c|c|c|c|c|c|c|}
\hline
$\alpha$&$D=5$&$D=6$&$D=7$&$D=8$&$D=9$&$D=10$&$D=11$\\
\hline
$0.00$&$0.7509-0.3639\imo$&$0.8127-0.5045\imo$&$0.8168-0.6111\imo$&$0.7923-0.6800\imo$&$0.7618-0.7173\imo$&&\\
$0.01$&$0.7489-0.3605\imo$&$0.8111-0.4950\imo$&$0.8198-0.5960\imo$&$0.8022-0.6633\imo$&$0.7770-0.7027\imo$&$0.7542-0.7237\imo$&$0.7367-0.7345\imo$\\
$0.02$&$0.7469-0.3572\imo$&$0.8092-0.4860\imo$&$0.8215-0.5816\imo$&$0.8096-0.6467\imo$&$0.7891-0.6874\imo$&$0.7688-0.7113\imo$&$0.7519-0.7250\imo$\\
$0.03$&$0.7449-0.3539\imo$&$0.8072-0.4773\imo$&$0.8224-0.5680\imo$&$0.8149-0.6310\imo$&$0.7985-0.6724\imo$&$0.7805-0.6987\imo$&$0.7642-0.7152\imo$\\
$0.04$&$0.7430-0.3507\imo$&$0.8049-0.4690\imo$&$0.8224-0.5552\imo$&$0.8187-0.6161\imo$&$0.8058-0.6582\imo$&$0.7898-0.6867\imo$&$0.7741-0.7059\imo$\\
$0.05$&$0.7410-0.3476\imo$&$0.8025-0.4611\imo$&$0.8219-0.5432\imo$&$0.8213-0.6023\imo$&$0.8113-0.6449\imo$&$0.7972-0.6754\imo$&$0.7821-0.6971\imo$\\
$0.06$&$0.7390-0.3446\imo$&$0.8001-0.4535\imo$&$0.8209-0.5320\imo$&$0.8230-0.5895\imo$&$0.8155-0.6326\imo$&$0.8032-0.6650\imo$&$0.7888-0.6890\imo$\\
$0.07$&$0.7370-0.3416\imo$&$0.7975-0.4463\imo$&$0.8197-0.5215\imo$&$0.8239-0.5777\imo$&$0.8187-0.6213\imo$&$0.8080-0.6554\imo$&$0.7943-0.6816\imo$\\
$0.08$&$0.7350-0.3386\imo$&$0.7949-0.4394\imo$&$0.8181-0.5117\imo$&$0.8244-0.5668\imo$&$0.8212-0.6109\imo$&$0.8120-0.6466\imo$&$0.7990-0.6748\imo$\\
$0.09$&$0.7330-0.3358\imo$&$0.7923-0.4328\imo$&$0.8164-0.5024\imo$&$0.8245-0.5566\imo$&$0.8231-0.6013\imo$&$0.8153-0.6385\imo$&$0.8030-0.6685\imo$\\
$0.10$&$0.7311-0.3329\imo$&$0.7896-0.4264\imo$&$0.8145-0.4937\imo$&$0.8242-0.5472\imo$&$0.8246-0.5925\imo$&$0.8180-0.6311\imo$&$0.8064-0.6628\imo$\\
$0.11$&$0.7291-0.3302\imo$&$0.7869-0.4204\imo$&$0.8126-0.4856\imo$&$0.8238-0.5384\imo$&$0.8257-0.5843\imo$&$0.8204-0.6242\imo$&$0.8094-0.6575\imo$\\
$0.12$&$0.7271-0.3274\imo$&$0.7842-0.4145\imo$&$0.8106-0.4778\imo$&$0.8232-0.5303\imo$&$0.8266-0.5768\imo$&$0.8223-0.6179\imo$&$0.8120-0.6526\imo$\\
$0.13$&$0.7252-0.3248\imo$&$0.7815-0.4089\imo$&$0.8085-0.4705\imo$&$0.8224-0.5226\imo$&$0.8272-0.5698\imo$&$0.8240-0.6120\imo$&$0.8143-0.6480\imo$\\
$0.14$&$0.7232-0.3222\imo$&$0.7788-0.4035\imo$&$0.8065-0.4636\imo$&$0.8216-0.5155\imo$&$0.8277-0.5632\imo$&$0.8255-0.6065\imo$&$0.8164-0.6438\imo$\\
$0.15$&$0.7213-0.3196\imo$&$0.7761-0.3984\imo$&$0.8044-0.4571\imo$&$0.8207-0.5088\imo$&$0.8280-0.5571\imo$&$0.8268-0.6014\imo$&$0.8182-0.6398\imo$\\
$0.16$&$0.7193-0.3171\imo$&$0.7735-0.3934\imo$&$0.8023-0.4509\imo$&$0.8198-0.5024\imo$&$0.8282-0.5513\imo$&$0.8279-0.5966\imo$&$0.8199-0.6361\imo$\\
$0.17$&$0.7174-0.3146\imo$&$0.7709-0.3885\imo$&$0.8002-0.4450\imo$&$0.8188-0.4965\imo$&$0.8284-0.5460\imo$&$0.8289-0.5921\imo$&$0.8214-0.6326\imo$\\
$0.18$&$0.7155-0.3122\imo$&$0.7683-0.3839\imo$&$0.7982-0.4393\imo$&$0.8178-0.4908\imo$&$0.8284-0.5409\imo$&$0.8297-0.5879\imo$&$0.8227-0.6294\imo$\\
$0.19$&$0.7136-0.3098\imo$&$0.7657-0.3794\imo$&$0.7962-0.4339\imo$&$0.8168-0.4855\imo$&$0.8284-0.5361\imo$&$0.8305-0.5839\imo$&$0.8239-0.6263\imo$\\
$0.20$&$0.7117-0.3075\imo$&$0.7632-0.3751\imo$&$0.7942-0.4288\imo$&$0.8158-0.4804\imo$&$0.8283-0.5316\imo$&$0.8311-0.5802\imo$&$0.8251-0.6234\imo$\\
\hline
\end{tabular}
\end{table}

\begin{table}
\caption{Fundamental quasi-normal modes of the Gauss-Bonnet black hole for the massless Dirac field localised on the $4D$-brane ($\kappa=1$). All quantities are measured in units of the radius of horizon. The first line ($\alpha=0$) data are taken from \cite{Kanti:2006ua}.}\label{tbl.s=0.5.k=1}
\begin{tabular}{|c|c|c|c|c|c|c|c|}
\hline
$\alpha$&$D=5$&$D=6$&$D=7$&$D=8$&$D=9$&$D=10$&$D=11$\\
\hline
$0.00$&$0.4413-0.3598\imo$&$0.4533-0.5105\imo$&$0.4193-0.6520\imo$&$0.3321-0.7862\imo$&$0.3144-1.2830\imo$&&\\
$0.01$&$0.4397-0.3557\imo$&$0.4530-0.4966\imo$&$0.4291-0.6201\imo$&$0.3848-0.7269\imo$&$0.2907-0.8213\imo$&$0.1457-0.9001\imo$&\\
$0.02$&$0.4386-0.3517\imo$&$0.4531-0.4834\imo$&$0.4366-0.5925\imo$&$0.4027-0.6820\imo$&$0.3480-0.7632\imo$&$0.2809-0.8266\imo$&$0.1867-0.8857\imo$\\
$0.03$&$0.4375-0.3479\imo$&$0.4528-0.4712\imo$&$0.4414-0.5691\imo$&$0.4163-0.6473\imo$&$0.3802-0.7156\imo$&$0.3361-0.7724\imo$&$0.2808-0.8255\imo$\\
$0.04$&$0.4364-0.3441\imo$&$0.4523-0.4600\imo$&$0.4444-0.5488\imo$&$0.4255-0.6190\imo$&$0.3988-0.6801\imo$&$0.3655-0.7334\imo$&$0.3238-0.7848\imo$\\
$0.05$&$0.4353-0.3405\imo$&$0.4516-0.4497\imo$&$0.4463-0.5312\imo$&$0.4314-0.5958\imo$&$0.4108-0.6524\imo$&$0.3840-0.7042\imo$&$0.3495-0.7552\imo$\\
$0.06$&$0.4342-0.3370\imo$&$0.4507-0.4402\imo$&$0.4475-0.5157\imo$&$0.4360-0.5760\imo$&$0.4193-0.6302\imo$&$0.3967-0.6814\imo$&$0.3668-0.7327\imo$\\
$0.07$&$0.4331-0.3336\imo$&$0.4497-0.4313\imo$&$0.4481-0.5019\imo$&$0.4393-0.5592\imo$&$0.4255-0.6119\imo$&$0.4060-0.6630\imo$&$0.3793-0.7149\imo$\\
$0.08$&$0.4320-0.3302\imo$&$0.4486-0.4230\imo$&$0.4484-0.4896\imo$&$0.4417-0.5446\imo$&$0.4304-0.5965\imo$&$0.4132-0.6478\imo$&$0.3889-0.7003\imo$\\
$0.09$&$0.4309-0.3270\imo$&$0.4475-0.4152\imo$&$0.4485-0.4784\imo$&$0.4436-0.5318\imo$&$0.4342-0.5832\imo$&$0.4190-0.6350\imo$&$0.3965-0.6883\imo$\\
$0.10$&$0.4298-0.3238\imo$&$0.4463-0.4079\imo$&$0.4484-0.4683\imo$&$0.4451-0.5205\imo$&$0.4373-0.5716\imo$&$0.4237-0.6240\imo$&$0.4027-0.6780\imo$\\
$0.11$&$0.4288-0.3207\imo$&$0.4451-0.4010\imo$&$0.4482-0.4590\imo$&$0.4463-0.5103\imo$&$0.4400-0.5617\imo$&$0.4277-0.6145\imo$&$0.4079-0.6692\imo$\\
$0.12$&$0.4277-0.3177\imo$&$0.4439-0.3945\imo$&$0.4479-0.4505\imo$&$0.4473-0.5012\imo$&$0.4423-0.5527\imo$&$0.4310-0.6061\imo$&$0.4123-0.6615\imo$\\
$0.13$&$0.4266-0.3148\imo$&$0.4428-0.3884\imo$&$0.4475-0.4426\imo$&$0.4481-0.4929\imo$&$0.4442-0.5447\imo$&$0.4340-0.5987\imo$&$0.4161-0.6548\imo$\\
$0.14$&$0.4255-0.3119\imo$&$0.4416-0.3825\imo$&$0.4471-0.4353\imo$&$0.4488-0.4853\imo$&$0.4459-0.5374\imo$&$0.4365-0.5920\imo$&$0.4194-0.6487\imo$\\
$0.15$&$0.4244-0.3092\imo$&$0.4404-0.3770\imo$&$0.4467-0.4285\imo$&$0.4493-0.4784\imo$&$0.4474-0.5308\imo$&$0.4388-0.5860\imo$&$0.4223-0.6433\imo$\\
$0.16$&$0.4234-0.3064\imo$&$0.4392-0.3717\imo$&$0.4463-0.4222\imo$&$0.4498-0.4720\imo$&$0.4487-0.5248\imo$&$0.4408-0.5806\imo$&$0.4249-0.6384\imo$\\
$0.17$&$0.4223-0.3038\imo$&$0.4380-0.3667\imo$&$0.4458-0.4162\imo$&$0.4503-0.4661\imo$&$0.4499-0.5193\imo$&$0.4426-0.5756\imo$&$0.4273-0.6340\imo$\\
$0.18$&$0.4212-0.3012\imo$&$0.4369-0.3619\imo$&$0.4453-0.4106\imo$&$0.4506-0.4606\imo$&$0.4510-0.5142\imo$&$0.4443-0.5710\imo$&$0.4294-0.6299\imo$\\
$0.19$&$0.4202-0.2986\imo$&$0.4358-0.3573\imo$&$0.4449-0.4054\imo$&$0.4509-0.4554\imo$&$0.4519-0.5095\imo$&$0.4458-0.5668\imo$&$0.4314-0.6262\imo$\\
$0.20$&$0.4191-0.2961\imo$&$0.4347-0.3529\imo$&$0.4444-0.4004\imo$&$0.4512-0.4506\imo$&$0.4528-0.5051\imo$&$0.4472-0.5629\imo$&$0.4331-0.6228\imo$\\
\hline
\end{tabular}
\end{table}

\begin{table}
\caption{Fundamental quasi-normal modes of the Gauss-Bonnet black hole for the gauge field localised on the $4D$-brane ($l=1$). All quantities are measured in units of the radius of horizon. The first line ($\alpha=0$) data are taken from \cite{Kanti:2006ua}.}\label{tbl.s=l=1}
\begin{tabular}{|c|c|c|c|c|c|c|c|}
\hline
$\alpha$&$D=5$&$D=6$&$D=7$&$D=8$&$D=9$&$D=10$&$D=11$\\
\hline
$0.00$&$0.5767-0.3175\imo$&$0.5840-0.4021\imo$&$0.5693-0.4490\imo$&$0.5527-0.4727\imo$&$0.5963-0.4845\imo$&&\\
$0.01$&$0.5761-0.3150\imo$&$0.5851-0.3970\imo$&$0.5729-0.4430\imo$&$0.5582-0.4672\imo$&$0.5462-0.4798\imo$&$0.5374-0.4866\imo$&$0.5310-0.4906\imo$\\
$0.02$&$0.5756-0.3125\imo$&$0.5860-0.3920\imo$&$0.5761-0.4370\imo$&$0.5631-0.4614\imo$&$0.5521-0.4748\imo$&$0.5435-0.4824\imo$&$0.5371-0.4871\imo$\\
$0.03$&$0.5750-0.3100\imo$&$0.5867-0.3870\imo$&$0.5789-0.4310\imo$&$0.5675-0.4557\imo$&$0.5572-0.4698\imo$&$0.5488-0.4783\imo$&$0.5422-0.4837\imo$\\
$0.04$&$0.5743-0.3076\imo$&$0.5873-0.3821\imo$&$0.5813-0.4251\imo$&$0.5712-0.4500\imo$&$0.5616-0.4649\imo$&$0.5533-0.4743\imo$&$0.5465-0.4806\imo$\\
$0.05$&$0.5737-0.3052\imo$&$0.5876-0.3773\imo$&$0.5833-0.4193\imo$&$0.5745-0.4444\imo$&$0.5654-0.4602\imo$&$0.5573-0.4705\imo$&$0.5501-0.4777\imo$\\
$0.06$&$0.5730-0.3028\imo$&$0.5879-0.3726\imo$&$0.5850-0.4137\imo$&$0.5773-0.4391\imo$&$0.5688-0.4556\imo$&$0.5606-0.4670\imo$&$0.5533-0.4749\imo$\\
$0.07$&$0.5723-0.3005\imo$&$0.5879-0.3680\imo$&$0.5864-0.4082\imo$&$0.5797-0.4339\imo$&$0.5717-0.4513\imo$&$0.5636-0.4636\imo$&$0.5560-0.4724\imo$\\
$0.08$&$0.5716-0.2982\imo$&$0.5879-0.3636\imo$&$0.5875-0.4029\imo$&$0.5818-0.4289\imo$&$0.5742-0.4472\imo$&$0.5661-0.4604\imo$&$0.5583-0.4701\imo$\\
$0.09$&$0.5709-0.2960\imo$&$0.5878-0.3592\imo$&$0.5884-0.3978\imo$&$0.5836-0.4241\imo$&$0.5764-0.4433\imo$&$0.5684-0.4575\imo$&$0.5604-0.4680\imo$\\
$0.10$&$0.5701-0.2937\imo$&$0.5875-0.3550\imo$&$0.5891-0.3929\imo$&$0.5851-0.4196\imo$&$0.5784-0.4396\imo$&$0.5704-0.4547\imo$&$0.5623-0.4660\imo$\\
$0.11$&$0.5694-0.2916\imo$&$0.5872-0.3509\imo$&$0.5897-0.3882\imo$&$0.5864-0.4152\imo$&$0.5801-0.4361\imo$&$0.5722-0.4521\imo$&$0.5640-0.4641\imo$\\
$0.12$&$0.5686-0.2894\imo$&$0.5867-0.3469\imo$&$0.5901-0.3837\imo$&$0.5876-0.4111\imo$&$0.5816-0.4328\imo$&$0.5738-0.4497\imo$&$0.5655-0.4624\imo$\\
$0.13$&$0.5678-0.2873\imo$&$0.5863-0.3430\imo$&$0.5904-0.3793\imo$&$0.5886-0.4072\imo$&$0.5830-0.4297\imo$&$0.5753-0.4474\imo$&$0.5668-0.4608\imo$\\
$0.14$&$0.5670-0.2852\imo$&$0.5857-0.3393\imo$&$0.5906-0.3751\imo$&$0.5894-0.4035\imo$&$0.5842-0.4267\imo$&$0.5766-0.4453\imo$&$0.5680-0.4593\imo$\\
$0.15$&$0.5662-0.2831\imo$&$0.5851-0.3356\imo$&$0.5907-0.3711\imo$&$0.5901-0.3999\imo$&$0.5853-0.4240\imo$&$0.5778-0.4432\imo$&$0.5692-0.4579\imo$\\
$0.16$&$0.5653-0.2810\imo$&$0.5845-0.3321\imo$&$0.5907-0.3672\imo$&$0.5908-0.3965\imo$&$0.5863-0.4213\imo$&$0.5789-0.4413\imo$&$0.5702-0.4566\imo$\\
$0.17$&$0.5645-0.2790\imo$&$0.5838-0.3286\imo$&$0.5906-0.3635\imo$&$0.5913-0.3932\imo$&$0.5872-0.4188\imo$&$0.5799-0.4395\imo$&$0.5711-0.4553\imo$\\
$0.18$&$0.5637-0.2771\imo$&$0.5830-0.3252\imo$&$0.5905-0.3599\imo$&$0.5917-0.3901\imo$&$0.5880-0.4164\imo$&$0.5808-0.4378\imo$&$0.5720-0.4541\imo$\\
$0.19$&$0.5628-0.2751\imo$&$0.5823-0.3220\imo$&$0.5903-0.3565\imo$&$0.5921-0.3872\imo$&$0.5887-0.4141\imo$&$0.5817-0.4361\imo$&$0.5728-0.4530\imo$\\
$0.20$&$0.5619-0.2732\imo$&$0.5815-0.3188\imo$&$0.5901-0.3532\imo$&$0.5925-0.3843\imo$&$0.5894-0.4119\imo$&$0.5824-0.4346\imo$&$0.5736-0.4519\imo$\\
\hline
\end{tabular}
\end{table}

\begin{figure}
\includegraphics[width=.5\textwidth,clip]{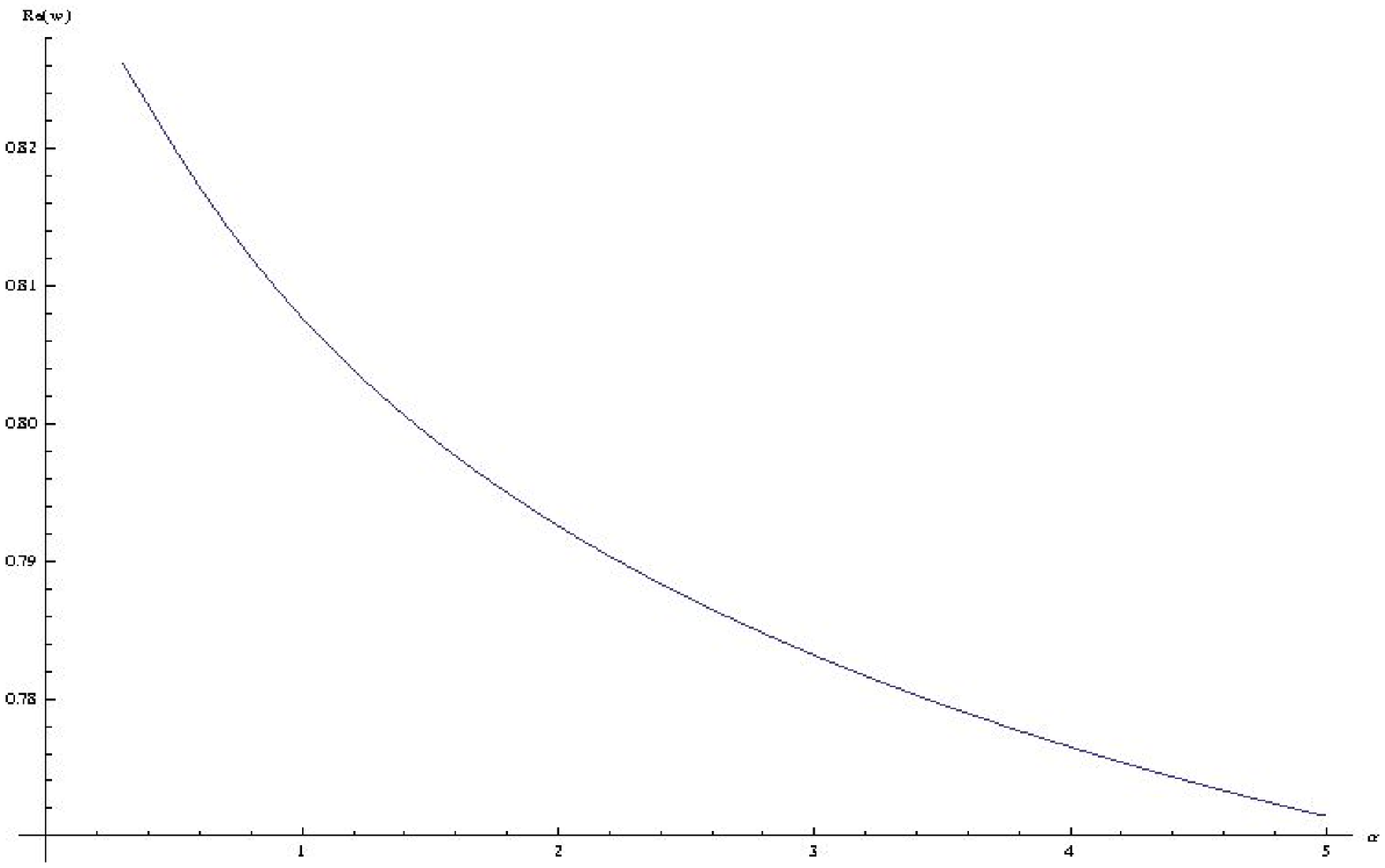}\includegraphics[width=.5\textwidth,clip]{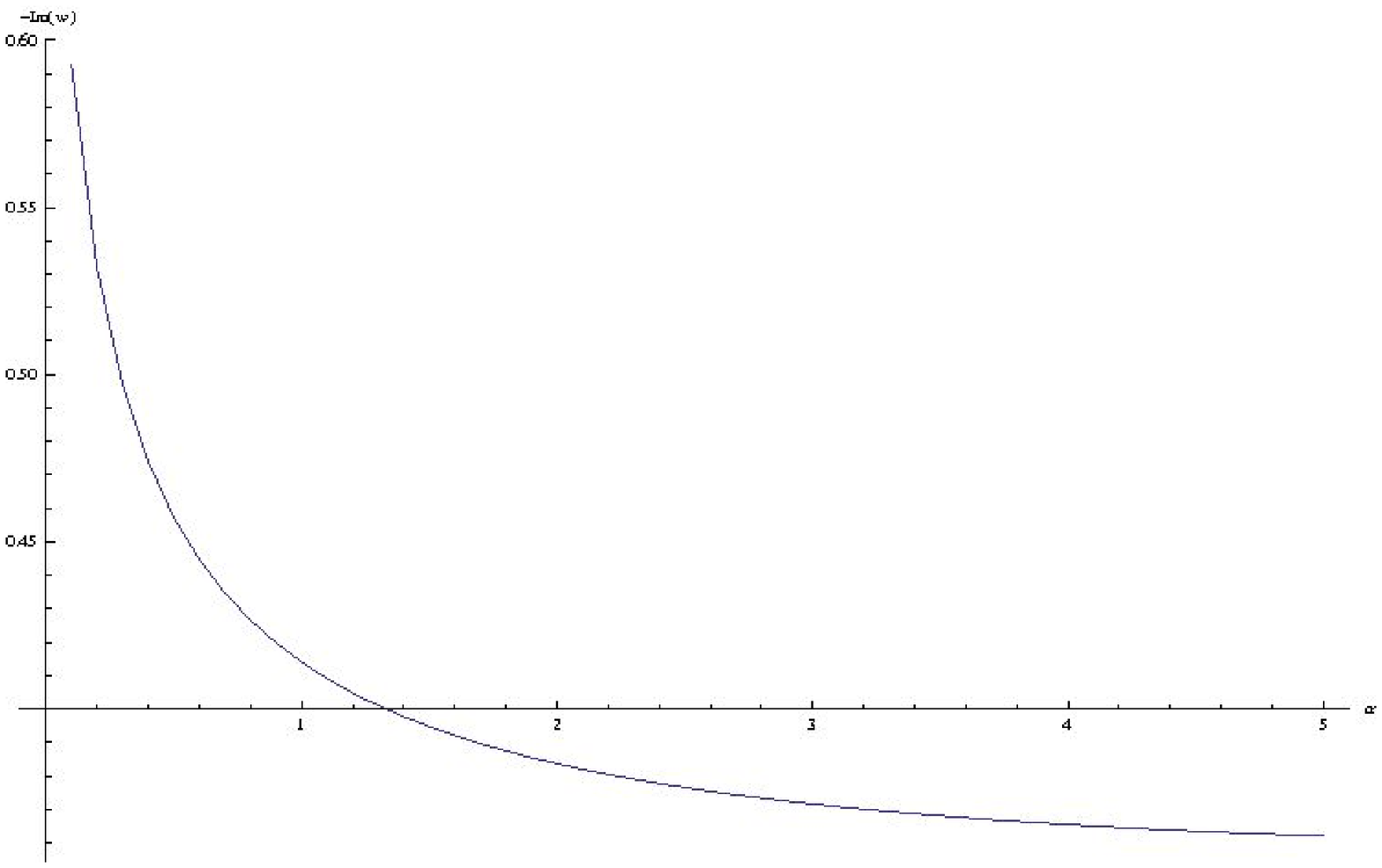}
\caption{Real and imaginary part of the dominant quasi-normal frequency of the scalar field localised on a $4D$-brane as a function of $\alpha$ ($D=9$, $l=1$).}\label{fig.D=9.s=0.l=1.w}
\end{figure}

We study the ringing of GB black hole using a numerical characteristic integration method \cite{Price-Pullin}, that uses the light-cone variables $u = t - r_\star$ and $v = t + r_\star$. In the characteristic initial value problem, initial data are specified on the two null surfaces $u = u_{0}$ and $v = v_{0}$. The discretization scheme we used, is
\begin{equation}\label{d-uv-eq}
R(N) = R(W) + R(E) - R(S) -\Delta^2\frac{V(W)R(W) + V(E)R(E)}{8} + \mathcal{O}(\Delta^4) \ ,
\end{equation}
where we have used the following definitions for the points: $N =(u + \Delta, v + \Delta)$, $W = (u + \Delta, v)$, $E = (u, v + \Delta)$ and $S = (u,v)$. The final \texttt{C++} programm that finds the time-domain profiles with arbitrary precision is available upon request.
\end{widetext}

\begin{figure}
\includegraphics[width=.45\textwidth,clip]{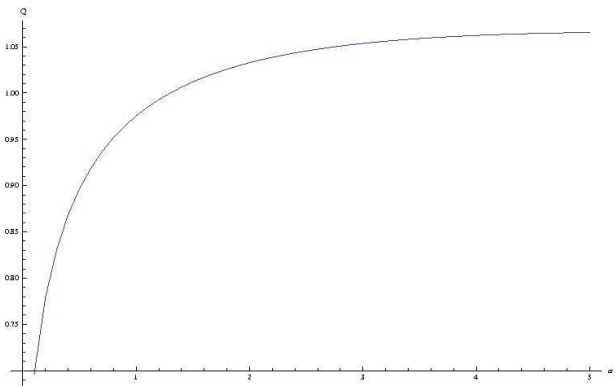}
\caption{The quality factor of the scalar field localised on a $4D$-brane as a function of $\alpha$ ($D=9$, $l=1$).}\label{fig.D=9.s=0.l=1.Q}
\end{figure}

\begin{figure}
\includegraphics[width=.45\textwidth,clip]{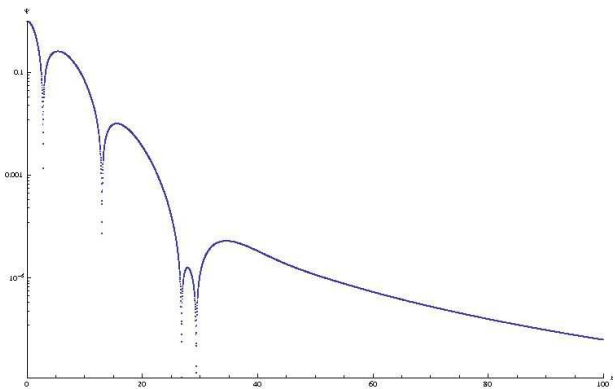}
\caption{The time-domain profile of the brane-localised massless scalar field ($D=5$, $\alpha=0.2$, $l=0$).}\label{fig.D=5.s=l=0.profile}
\end{figure}

\begin{figure}
\includegraphics[width=.45\textwidth,clip]{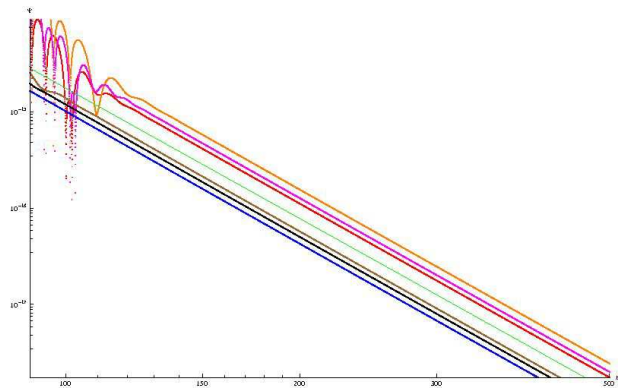}
\caption{The late-time behavior of the scalar, gauge and Dirac field localised on a $4D$-brane ($D=8$, $l(\kappa)=1$) for $\alpha=0.5$ (blue, brown, black respectively) (lower lines) and $\alpha=5.0$ (red, orange, magenta respectively) (upper lines) together with a function $\propto t^{-9}$ (green) (middle line).}\label{fig.D=8.l=1.latetime}
\end{figure}

\begin{figure}
\includegraphics[width=.45\textwidth,clip]{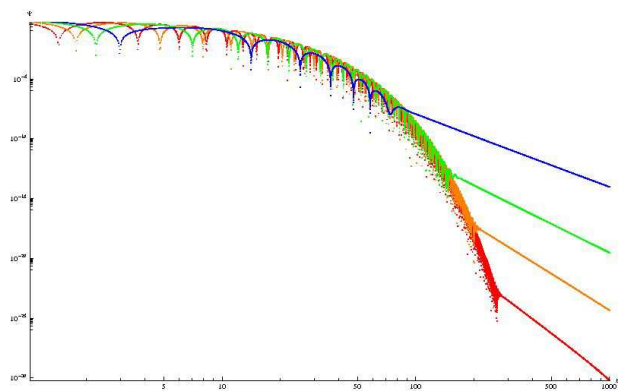}
\caption{The time-domain profiles of the brane-localised massless scalar field ($D=7$, $\alpha=5$) for $l=0$ (blue), $l=1$ (green), $l=2$ (orange), $l=4$ (red). The larger $l$ corresponds to the longer life of quasi-normal ringing and the quicker tail decay.}\label{fig.D=7.a=5.0.scalar.profiles}
\end{figure}

\begin{table}
\caption{Quasi-normal modes of the $7D$ Gauss-Bonnet black hole for the test scalar field localised on a $4D$-brane ($l=0$). Two dominant modes ($\omega_0$ and $\omega_1$) reach almost the same imaginary part for $\alpha=[0.07..0.10]$.}\label{tbl.D=7.s=l=0}
\begin{tabular}{|c|c|c|}
\hline
$\alpha$&$\omega_0$&$\omega_1$\\
\hline
$0.01$&$0.38-0.0638\imo$&$0.75-0.3647\imo$\\
$0.02$&$0.34-0.0824\imo$&$0.75-0.3698\imo$\\
$0.03$&$0.32-0.1222\imo$&$0.76-0.3808\imo$\\
$0.04$&$0.32-0.2010\imo$&$0.78-0.4153\imo$\\
$0.05$&$0.29-0.3197\imo$&$0.79-0.5038\imo$\\
$0.06$&$0.21-0.4835\imo$&$0.71-0.6089\imo$\\
\hline
$0.11$&$0.32-0.4794\imo$&$0.62-0.9972\imo$\\
$0.12$&$0.33-0.4344\imo$&$0.74-1.0199\imo$\\
$0.13$&$0.34-0.4065\imo$&$0.82-1.0253\imo$\\
$0.14$&$0.34-0.3871\imo$&$0.89-1.0236\imo$\\
$0.15$&$0.35-0.3726\imo$&$0.94-1.0189\imo$\\
$0.16$&$0.35-0.3615\imo$&$0.99-1.0130\imo$\\
$0.17$&$0.35-0.3526\imo$&$1.03-1.0065\imo$\\
\hline
\end{tabular}
\end{table}

We decompose the perturbation evolution profile into a set of damping oscillation
$$R(t,r)=\sum_{\omega}A_r e^{-\imo\omega t},$$
so that negative imaginary part of the complex frequency $\omega$ corresponds to the damping mode.

In order to check that the numerical accuracy is good enough, we repeated the integration procedure with smaller $\Delta$ and higher numerical precision. As a result, we obtained the same picture of the perturbation evolution, which means that the result is correct within required precision. The accuracy of the QNMs values we present is limited by finite time of the quasi-normal epoch, rather than precision of the numerical integration. Another way to check our calculation is going to the limit of $\alpha=0$ in which we must get the Schwarzschild black hole projected on a brane. From the tables \ref{tbl.s=0.l=1}, \ref{tbl.s=0.5.k=1}, \ref{tbl.s=l=1} one can find that the first line values, taken from \cite{Kanti:2006ua}, are close to our results for small $\alpha$.

We see that the oscillations decay faster for higher $D$ as well as it was found for the pure Schwarzschild black hole \cite{Kanti:2005xa}. The Gauss-Bonnet term causes the perturbations to decay slower. Note, that the same effect takes place for the $D$-dimensional scalar field \cite{Abdalla:2005hu} and for the gravitational perturbations \cite{Konoplya:2008ix} of the $D$-dimensional Gauss-Bonnet black hole. The real part of the quasi-normal frequencies has a more complicated behavior: for $D=5$ it decreases as $\alpha$ grows, but for higher dimension cases it starts growing first and then decreases after some value of $\alpha$ is reached. This value of $\alpha$ depends on $D$ and the perturbed field.

Since the Gauss-Bonnet solution supposes $\alpha$ to be small, we considered mainly the region $\alpha\leq0.2$, which is suggested by the limit of stability of the Gauss-Bonnet black hole in the $5$-dimensional space-time. Yet, figure \ref{fig.D=9.s=0.l=1.w} represents qualitatively the fundamental frequency behavior for larger $\alpha$.

Despite the real part changes its behavior, the quality factor $Q=\frac{1}{2}\left|\frac{Re(\omega)}{Im(\omega)}\right|$ increases as $\alpha$ grows for all fields and all values of $D$. This takes place also for larger $\alpha$ (see FIG. \ref{fig.D=9.s=0.l=1.Q}).

The spherically symmetric perturbation of a scalar field localised on a $4D$-brane is more difficult to study, because the time of the quasi-normal ringing epoch happens to be of the dominant oscillation period order (see FIG. \ref{fig.D=5.s=l=0.profile}). That is why we were unable to calculate the oscillation frequency with a good accuracy. Also, we cannot distinguish the modes that decay with almost the same rate. Yet, we can conclude that for some $\alpha$ two dominant oscillations reach almost the same imaginary part (see Table \ref{tbl.D=7.s=l=0}). The same feature has been recently observed for the scalar type of the gravitational perturbations of the $D$-dimensional Gauss-Bonnet black hole \cite{Konoplya:2008ix}.

As for the ordinary fields in higher dimensional space-time, the quasi-normal ringing epoch is followed by the asymptotical tails with power-law decay with time. From the figure \ref{fig.D=8.l=1.latetime} we can see that the decay law is the same for the tails of different field perturbations and different $\alpha$. These tails are universal and depend only on $D$ and $l$. The higher $D$ or $l$ is the quicker decay of the tails we observe. We found some time-domain profiles for different $D$ and $l$ (see for instance FIG. \ref{fig.D=7.a=5.0.scalar.profiles}) and conclude that their decay law is $$\Psi\propto t^{1-D-2l},$$ which is different from that for fields in higher dimensional space-time \cite{Abdalla:2005hu} and can be described by the same analytical formula for odd and even values of $D$.

\section{Quasi-normal modes in the large multipole number limit}\label{sec:lmlimit}
Using the WKB formula \cite{WKB} one can find the large multipole limit for the QNMs of electromagnetic or scalar field perturbations
\begin{eqnarray}\label{largemultipole}
&&\omega = \frac{\sqrt{f(R)}}{R}\left(l+\frac{1}{2}\right)-\\\nonumber&&-\imo\sqrt{3f(R)\frac{d}{dR}\left(\frac{f'(R)}{R}\right)-\frac{1}{2}\frac{d(f(R)f'(R))}{dR}}\left(n+\frac{1}{2}\right) +\\\nonumber &&+ {\cal O}\left(\frac{1}{l}\right),\qquad n=0,1,2\ldots
\end{eqnarray}
where $R$ is the point where the effective potential reaches its maximum when $l\gg1$. $R$ satisfies the equation
\begin{equation}\label{Req}
Rf'(R)=2f(R).
\end{equation}

For small $\alpha$ $R$ can be expanded as
\begin{equation}\label{Rexp}
R=R_0+\alpha R_1+\ldots
\end{equation}

Substituting (\ref{Rexp}) into (\ref{Req}) we find the coefficients
\begin{eqnarray}&&\left(\frac{R_0}{r_0}\right)^{D-3}=\frac{D-1}{2},\nonumber\\
&&R_1=\frac{D-4}{D-1}\frac{R_0^2(D-1)-4r_0^2}{2R_0r_0^2}\ldots\nonumber
\end{eqnarray}

Then (\ref{largemultipole}) reads
\begin{eqnarray}
\omega&=&\Omega_R(1+\alpha A_1+\ldots)\left(l+\frac{1}{2}\right)-\\\nonumber&&-\imo\Omega_I\left((1+\alpha B_1+\ldots\right)\left(n+\frac{1}{2}\right)+ {\cal O}\left(\frac{1}{l}\right).
\end{eqnarray}
For perturbations of Dirac field we obtain
\begin{eqnarray}
\omega&=&\Omega_R(1+\alpha A_1+\ldots)\kappa_\pm-\\\nonumber&&-\imo\Omega_I\left((1+\alpha B_1+\ldots\right)\left(n+\frac{1}{2}\right)+ {\cal O}\left(\frac{1}{\kappa_\pm}\right).
\end{eqnarray}

After some algebra we find \cite{Kanti:2005xa}
$$\Omega_R=\frac{1}{R_0}\sqrt{\frac{D-3}{D-1}},\quad\Omega_I=\frac{1}{R_0}\frac{D-3}{\sqrt{D-1}}.$$
Therefore, as it usually happens for higher dimensional black holes, the quality factor decreases with $D$
$$Q\sim\frac{\Omega_R}{\Omega_I}=\frac{1}{\sqrt{D-3}}.$$

The corrections of the first order of $\alpha$ are given by
\begin{eqnarray}\nonumber
A_1&=&-\frac{1}{r_0^2}\frac{D-4}{D-1}\left(\frac{D-1}{2}-\frac{r_0^2}{R_0^2}\right)<0,\\\nonumber
B_1&=&-\frac{1}{r_0^2}\frac{D-4}{D-1}\left(\frac{D-1}{2}+(D-2)\frac{R_0^2}{r_0^2}\right)<0.
\end{eqnarray}
Thus we conclude, that, at least for large multipole numbers, the real and the imaginary parts of $\omega$ decrease their absolute values with $\alpha$ for sufficiently small $\alpha$. We see, thereby, that for large multipoles the dependance of the imaginary part on $\alpha$ is basically the same as for small ones. Since $B_1\ll A_1$ for $D\gg1$ the real part of $\omega$ decreases much slower. We have seen that for small $l$ this effect is suppressed by the correction of ${\cal O}(l^{-1})$, implying a complicated dependence on $\alpha$.

\section{Conclusion}\label{sec:conclusion}
We have studied the evolution of massless fields localised on a $4D$-brane in the exterior of $D$-dimensional black holes in Gauss-Bonnet theory for $D=5..11$. We have found the dependance of the quasi-normal frequency on the Gauss-Bonnet coupling parameter $\alpha$. We conclude that, if we take into account $\alpha$, the damping rate is decreased and, thereby, the black hole appears to be better oscillator. This effect could be significant for small black holes, giving us a possible opportunity to evaluate the number of extra dimensions and the Gauss-Bonnet parameter if we were able to detect quasi-normal ringing from mini black holes, produced by LHC.

The asymptotically late time tails do not depend on the field spin and the Gauss-Bonnet parameter. They decay according to the power law, which depends only on the number of extra dimensions and the multipole number.

\begin{acknowledgments}
GNU Multiple Precision Arithmetic Library was used to obtain the presented results.\\
This work was supported by \emph{Funda\c{c}\~ao de Amparo \`a Pesquisa do Estado de S\~ao Paulo (FAPESP)}, Brazil.\\
I would like to thank Roman Konoplya for proposing this problem to me and reading the manuscript.

\end{acknowledgments}

\end{document}